\documentclass[12pt]{article}
\usepackage{amsmath,amssymb,graphicx,epic}

\renewcommand{\a}{\alpha}
\renewcommand{\b}{\beta}

\renewcommand{\d}{\delta}

\renewcommand{\L}{\Lambda}
\newcommand{\LL}{\rm L}

\newcommand{\s}{\sigma}

\renewcommand{\P}{\Phi}

\renewcommand{\v}{\varphi}

\newcommand{\pa}{\partial}

\newcommand{\be}{\begin{equation}}
\newcommand{\ee}{\end{equation}}
\newcommand{\beq}{\begin{eqnarray}}
\newcommand{\eeq}{\end{eqnarray}}
\newcommand{\nn}{\nonumber\\}

\newcommand{\cl}{{\cal L}}

\newcommand{\cv}{{\cal V}}

\newcommand{\E}{E_{10}}
\newcommand{\K}{K(E_{10})}
\newcommand{\KE}{\E/\K}
\newcommand{\lag}{\mathfrak{g}}
\newcommand{\Ks}{K_{\rm{string}}}

\begin{document}
\begin{flushright}
{\small AEI-2006-023\\IHES/P/06/21\\ MIT-CTP-3735 \\  ULB-TH/06-07} \\
{\tt hep-th/0604143}\\[15mm]
\end{flushright}

\begin{center}
{\Large \sc Curvature Corrections and Kac--Moody Compatibility
  Conditions}\\[2cm]

Thibault Damour${}^{a}$, Amihay Hanany${}^{b}$, Marc
Henneaux${}^{c,d}$,\\ Axel Kleinschmidt${}^{e}$ and Hermann
Nicolai${}^{e}$

\footnotesize \vspace{.5 cm}

${}^a${\em Institut des Hautes Etudes Scientifiques, \\
35, route de Chartres, F-91440 Bures-sur-Yvette, France}

\vspace{.2cm}

${}^b${\em Center for Theoretical Physics,\\
Massachusetts Institute of Technology,\\
Cambridge MA 02139-4307, USA}
\vspace{.2cm}

${}^c${\em Universit\'e Libre de Bruxelles and International
Solvay Institutes,
\\ULB-Campus Plaine
C.P. 231\\ Boulevard du Triomphe, B-1050 Bruxelles, Belgium}\\

\vspace{.2cm}

${}^d${\em Centro de Estudios
Cient\'{\i}ficos (CECS), Casilla 1469, Valdivia, Chile}\\

\vspace{.2cm}

${}^e${\em Max-Planck-Institut f\"ur Gravitationsphysik
(Albert-Einstein-Institut)\\
M\"uhlenberg 1, D-14476 Potsdam, Germany}\\

\vspace{1cm}

\begin{tabular}{p{12cm}}
\hspace{5mm}{\bf Abstract:} We study possible restrictions on the
structure of curvature corrections to gravitational theories in
the context of their corresponding Kac--Moody algebras, following
the initial work on $E_{10}$ in Class. Quant. Grav.  {\bf 22}
(2005) 2849. We first emphasize that the leading quantum
corrections of M-theory can be naturally interpreted in terms of
{\em (non-gravity) fundamental weights} of $E_{10}$. We then
heuristically explore the extent to which this remark can be
generalized to all over-extended algebras by determining which
curvature corrections are compatible with their weight structure,
and by comparing these curvature terms with known results on the
quantum corrections for the corresponding gravitational theories.
\end{tabular}
\end{center}

\break

\section{Introduction}

The study of the BKL limits \cite{BKL} of coupled gravity-matter
systems near a space-like singularity has revealed an interesting
`correspondence' between the emerging cosmological billiard and over-extended
Kac--Moody algebras (KMAs). The cosmological billiard describes
the dynamics of a few effective degrees of freedom (spatial Kasner
exponents and dilatons) moving in a space bounded by walls that
become sharper as one goes towards the singularity. The unexpected
connection to the theory of KMAs is that the position of these
walls is related to root vectors in the root lattice of certain
special KMAs called the over-extended algebras
\cite{DaHe01,DadeBuHeScho02} (for a review and references see
\cite{DaHeNi03}). Important aspects of the classical
two-derivative action are reflected in algebraic properties of the
corresponding KMAs. For example, regularity or chaos of the motion as one
approaches the singularity is tied to hyperbolicity of the algebra
\cite{DHJN}: for hyperbolic KMAs there is chaos.

In recent work devoted to M-theory \cite{DaNi05}, it was shown
that the higher-derivative quantum corrections to the action
admit an interpretation in terms
of the Kac--Moody structure. More precisely, higher-order corrections in
the curvature (Riemann or Weyl) tensor and the 4-form field
strength were associated in \cite{DaNi05} to certain negative {\em
imaginary} roots on the $\E$ root lattice. There are two reasons
for the potential importance of this result, namely $(i)$ the
possibility that the `geodesic' $\KE$ $\s$-model may contain
hidden information about (perturbative) higher order corrections
of M-Theory to arbitrary orders, and $(ii)$ the fact that this
result may  allow one to understand the physical significance of
imaginary roots (recall that, also on the mathematical side, this
is where the main obstacles towards a better understanding
of indefinite KMAs lie).

Here, we elaborate on these results and conjecturally generalize them to other
over-extensions of finite-dimensional simple Lie
algebras.\footnote{The particular case of over-extended $G_2$ was already
analysed in \cite{MiMoYa05}.} For simplicity, we only consider the
pure curvature corrections, which provide the dominant terms; the
generalization of our results to other types of fields ($p$-forms)
is straightforward, at least in principle.\footnote{Corrections
involving field strengths give rise to billiard walls which are
`hidden' behind those from curvature corrections \cite{DaNi05}.}

We first observe that the imaginary roots that describe the
quantum corrections in the $\E$ case are not just any arbitrary
roots  but, rather, are the {\em dominant non-gravity weights}.
More precisely, as shown in section \ref{E10} below, the root of
$\E$ associated with the leading terms in the first quantum
corrections $R^4$ quartic in the curvature turns out to be the
fundamental weight $\Lambda_{10}$ conjugate to the `non-gravity'
root $\alpha_{10}$, i.e., to the (`exceptional') root which does
not lie on the $A_9 \equiv
\mathfrak{sl}(10)$ `gravity line' (see figure 1).  This property
ensures that $\Lambda_{10}$ is invariant under permutations of the
spatial directions (Weyl group of $\mathfrak{sl}(10)$). The
leading terms in the expected subsequent quantum corrections $R^7$,
$R^{10}$ {\it etc.} are associated with positive integer multiples of
$\Lambda_{10}$ and are thus also invariant under the Weyl group of
$\mathfrak{sl}(10)$.

\begin{figure}
\begin{center}
\scalebox{1}{
\begin{picture}(340,60)
\put(5,-5){$1$} \put(45,-5){$2$} \put(85,-5){$3$}
\put(125,-5){$4$} \put(165,-5){$5$} \put(205,-5){$6$}
\put(245,-5){$7$} \put(285,-5){$8$} \put(325,-5){$9$}
\put(260,45){$10$} \thicklines
\multiput(10,10)(40,0){9}{\circle{10}}
\multiput(15,10)(40,0){8}{\line(1,0){30}}
\put(250,50){\circle{10}} \put(250,15){\line(0,1){30}}
\end{picture}}
\caption{\label{e10dynk}\sl Dynkin diagram of $\E$ with numbering
of nodes.}
\end{center}
\end{figure}

This suggests that it might not be the root lattice which is relevant
for the correction terms, but rather the {\em weight lattice}. We recall
that the weight lattice is the lattice spanned by the fundamental weights
$\L_i$ obeying
\be\label{funwts}
\frac{2(\L_i|\a_j)}{(\a_i|\a_i)} = \d_{ij},
\ee
where the $\a_i$ are the simple roots spanning the root lattice. The
weight lattice always contains the root lattice as a sublattice
but is generically finer. We also recall that the set of
`dominant weights' is defined by taking integral linear combinations of the
fundamental weights with {\em non-negative} coefficients.
Correspondingly, we define
`dominant non-gravity weights' by taking
 non-negative integral linear combinations of those
fundamental weights not conjugate to roots on the `gravity line'.
Here, we shall generally define the `gravity line' by
 the simple roots associated with the so-called
`symmetry walls', i.e., the Hamiltonian contributions
which are related to the off-diagonal components
of the metric \cite{DaHeNi03}.

The purpose of this paper is to explore to what extent this
`botanical observation' inspired by the M-theory/$E_{10}$ analysis
(concerning the role of {\em dominant non-gravity weights}
of the corresponding over-extended KMA),
can be extended to higher-order curvature corrections
in other (super)gravity models. We shall therefore systematically
determine, for all over-extended Kac--Moody algebras
which higher order curvature
corrections are associated with dominant non-gravity weights, and
compare the results of this heuristic `algebraic selection rule'
to the currently known quantum corrections for the corresponding
gravitational theories.

{}For $\E$, the root and the weight lattices happen to coincide,
and therefore the
distinction between weights and roots did not play any role in the
analysis of~\cite{DaNi05}. In fact, $\E$ is the {\em only}
over-extended KMA for which the root lattice (usually designated
by $Q(\E)$) has this property: $Q(\E)=$ II$_{9,1}$ is the unique
even self-dual Lorentzian lattice in ten dimensions, and such
lattices are known to exist only in $2 +8n$ dimensions \cite{CS}.

{}For other algebras, however, we will find that there are correction
terms associated
with genuine weights {\em outside} the root lattice.  This fact
appears to imply that these corrections cannot be described within
the geodesic $\s$-model, and that one will have to augment the
$\s$-model Lagrangian by additional terms related to these weights
if the correspondence is to be extended to higher order corrections.
{\em Put differently, it is only for the maximally extended $\E$
model that one can argue, as was done in \cite{DaNi05}, that not
only the effective low energy theory, but the entire tower of higher
order corrections might be understood on the basis of a single
geodesic $\s$-model Lagrangian.}

Even in the $\E$ case, the observation that the imaginary roots that
arise are actually fundamental weights sheds a new, potentially
interesting, light on the structure of the subleading quantum
corrections since fundamental weights are naturally linked with
representations.  We indicate that the pattern for these subleading
terms uncovered in \cite{DaNi05} is indeed similar to that of a
lowest-weight representation (albeit a non-integrable one).

The assumption explored here that quantum corrections are controlled
by dominant
non-gravity weights gives us restrictions on the type of curvature
corrections which are consistent with the algebraic structure. We
have analysed these restrictions for all over-extended algebras
$\lag^{++}$ in split real form. The resulting restrictions are quite
satisfactory for M-theory~-- as originally found in \cite{DaNi05}~--
but there are mismatches which remain to be understood for the other
string-related cases.  This is particular striking for types IIA and
IIB, where some corrections known to appear do not define points on
the weight lattice.  Although this can be blamed on the singular
field theory limits involved in passing from M-theory to the
ten-dimensional models, we lack a deeper group-theoretical reason as
to why one finds perfect matching in some cases and not in
others.

Our paper is organized as follows. We first recall how to compute the
walls for a given Lagrangian and powers of the curvature tensor in
section~\ref{wallsec}. Section~\ref{E10} re-examines the case of
eleven-dimensional gravity and $\E$ using the dominant weight
perspective. In section~\ref{DBE10} we then repeat the analysis for
$D\E$ and $B\E$ for which the introduction of the weight lattice
becomes evident. The analogous results for the remaining over-extended
algebras are presented in section~\ref{g++}, illustrating the
particular properties of the ten-dimensional hyperbolic KMAs in the
class of all over-extended algebras. In section~\ref{EAB}, we analyse
$\E$ under decompositions appropriate for interpretations in terms of
the ten-dimensional IIA and IIB theories. Concluding remarks can be
found in section~\ref{concl}.

\section{Weights and curvature monomials}
\label{wallsec}

We follow the systematics of \cite{DaHeNi03,DadeBuHeScho02} for
constructing walls corresponding to any term in an effective
Lagrangian of the form
\be\label{genlag}
\cl = \cl^{(0)} + \sum_{m\ge1} ({\LL}^s)^m \cl^{(m)},
\ee
where $\cl^{(0)}$ is the lowest order Lagrangian (quadratic in
derivatives), and ${\LL}$ is a dimensionful expansion parameter
of dimension [Length] for the higher derivative correction
terms $\cl^{(m\ge 1)}$. The integer $s$ depends on the theory
in question; for instance, in the case of string theories one has
$s=2$ and $\LL^2= \a'$. We will take
$\cl^{(0)}$ to be the Lagrangian of the maximally oxidized theory
for a simple split Lie algebra $\lag$. These Lagrangians were
studied and given in \cite{CrJuLuPo99}; in particular, the maximal
space-time dimension $D=(d+1)$ was determined for all $\lag$. The
wall forms are already completely determined by $\cl^{(0)}$
\cite{DaHeNi03,DadeBuHeScho02}. They are parametrised by
a dilaton field $\v$ (when it exists in dimension $D$),
and by the
logarithmic {\em scale factors} $\b^a$ (for $a=1,\ldots, d$),
which appear via the following split of the spatial vielbein
\be
e_m{}^a = \exp(-\b^a) \theta_m{}^a
\ee
where the spatial frame $\theta_m{}^a$  `freezes' near the singularity
\cite{DaHeNi03}. 
These wall forms are explicitly written as
\be \label{wallform0}
w(\b,\v) = \sum_{a=1}^d p_a \b^a + p_\v \v.
\ee
The inner product between two such
wall forms $w$ and $w'$ is determined from the Hamiltonian constraint
following from the Einstein-Hilbert action in the standard way,
and given by~\footnote{Our conventions here differ from \cite{DadeBuHeScho02}
by a factor of $2$ for the dilaton terms since we normalise the
dilaton kinetic term with a factor of $1/2$ in the Einstein frame.
When there are several dilaton fields $\vec \v \equiv (\v_1,\v_2, \cdots)$,
the dilaton contributions in the wall forms and the wall scalar product
read $\vec p_\v \cdot \vec\v$ and $2\vec p_\v \cdot \vec p'_\v$, respectively.}
\be
(w|w') = \sum_{a=1}^d p_a p'_a - \frac1{d-1}
\left(\sum_{a=1}^d p_a \right) \left(\sum_{a=1}^d p'_a \right)
    + 2 p_\v  p'_\v.
\ee
For KMAs with non-symmetric generalized Cartan matrix this will yield
the symmetrized form with some values different from $2$ on the
diagonal since not all roots have the same length. For later use, we
also define the logarithm of the spatial volume factor
$\det e = \sqrt{g}$, namely
\be
\s = \sum_{a=1}^d \b^a.
\ee

Near a space-like singularity, the Lagrangian~(\ref{genlag}) can be
replaced by an effective Lagrangian describing the dynamics of the
logarithmic scale factors $\b^a$ and the dilaton $\v$ in an effective
potential $\cv$ that behaves asymptotically like a
sum of (exponentially sharp) billiard walls
\be
\cv(\b,\v) \sim \sum_w c_w e^{-2w(\b,\v)}
\ee
for a number of walls $w$. In all cases, the {\em dominant walls}
(those which have the largest contribution in the limit) are
located at positions describing the fundamental Weyl chamber of the
Kac--Moody-theoretic over-extension $\lag^{++}$ of the symmetry
algebra $\lag$ \cite{DaHeNi03}.

{}For simplicity, we will be concerned with curvature correction
terms of the type~\footnote{Here, $R^N$ denotes some $N$th- order
polynomial in the curvature tensor. Note that, while the inclusion
of Ricci and scalar curvature terms do not matter for the leading
terms, they do  for the subleading ones~\cite{DaNi05}.} (with
$s=2$ and $\LL^2 = \a'$) 
\be 
\cl^{(N-1)} \sim \sqrt{-G}\, R^N e^{K \v}. 
\ee 
As was shown in \cite{DaNi05} the dominant contribution
of such a term to the Hamiltonian density scales as 
\be
e^{2(N-1)\s +K \v} 
\ee 
in the asymptotic limit. Equating this with
a Hamiltonian wall potential $e^{-2w(\b,\v)}$ yields the
corresponding wall form 
\be\label{wallform} 
w_{N,K} (\b,\v)= -(N-1)\s -\frac12 K \v. 
\ee 
(or $w_{N,K} (\b,\vec\v)= -(N-1)\s
-\frac12 \vec{K}\cdot\vec\v$ in the multi-dilaton case). It is
important to keep in mind that (\ref{wallform}) is the form of the
wall {\em in the Einstein frame}. In order to make contact with
known results from string perturbation theory it will be necessary
to convert such a wall into the string frame. The notion of
Einstein frame exists in any dimension, and means that the
Einstein Hilbert term appears without dilatonic prefactors. By
contrast, the notion of string frame most readily applies in
$D=10$ dimensions, and means that the tensor-scalar part of the
action reads $e^{-2\P} \sqrt{-\tilde{G}}( \tilde{R} + 4
(\tilde{\pa} \P)^2)$,  where $\P$ is the standard
string dilaton field (with $g_s = e^{\P}$).
This parametrisation is natural if there is an
underlying world-sheet theory in which $\P$ couples to the
world-sheet curvature scalar. The effective theory will then admit an
expansion in genera of the world-sheet and higher order corrections at
$g$-loop order in string theory come with a factor
$e^{(2g-2)\P}$.

In {\em ten dimensions} the relation between the two frames, and
the two dilatons, as obtained by identifying\footnote{Where the
tilde on $\partial$ indicates that the partial derivatives are to
be contracted with the metric $\tilde{G}^{MN}$.} 
\be
\label{tensorscalar} \sqrt{-G}\big[ R - \frac12 (\pa \v)^2 \big] =
e^{-2\P} \sqrt{-\tilde{G}}\big[ \tilde{R} +  4 (\tilde{\pa} \P)^2 \big] 
\ee 
reads 
\be\label{10} 
\tilde{G}_{MN} = e^{\frac12\P} G_{MN} \qquad;\quad \P = \v \, , 
\ee 
where $G_{MN}$ and
$\tilde{G}_{MN}$ denote the metrics in the Einstein frame and the
string frame, respectively. The corresponding relation for the
higher order terms is therefore \be \sqrt{-G} R^N =  e^{\frac12
(N-5)\v} \sqrt{-\tilde{G}} \tilde{R}^N + \ldots \ee {}From this we
immediately read off the formula relating the coefficient $K$
multiplying the dilaton for the corresponding wall forms between
the two frames in ten dimensions, which reads
\be\label{eintostring} K\rightarrow \Ks\equiv K+\frac12(N-5). \ee
In our survey of KMAs we will also make use of the generalization
of this formula to $D\neq 10$ dimensions. Here, we define the
string frame in any dimension $D$ by requiring that the
tensor-scalar sector of the action density reads as the right-hand
side of equation (\ref{tensorscalar}) above. This yields  \be
\label{D} \tilde{G}_{MN} = e^{\frac{4}{D-2}\P} G_{MN} \qquad ;
\quad \P = \sqrt{\frac{D-2}{8}}\v \, , \ee instead of the $D=10$
result (\ref{10}) above. The corresponding generalization of the
transformation law for the dilaton coupling coefficient reads
\be\label{eintostring1} \Ks\equiv \sqrt{\frac{8}{D-2}}\,
K+\frac2{D-2}(2N-D). \ee

\section{$\E$ revisited}
\label{E10}

\subsection{Non-gravity weight $\L_{10}$ and $R^4$ correction}
We briefly recall the result of \cite{DaNi05} for $D=11$
supergravity and $\E\equiv E_8^{++}$, for which there is no dilaton
present. The dominant walls (=simple roots) are given explicitly by
\begin{align}
\a_1
&= (0,0,0,0,0,0,0,0,-1,1)& \Rightarrow &&\a_1(\b) &= -\b^9+\b^{10},&
\nn \vdots\nn \a_9 &=
(-1,1,0,0,0,0,0,0,0,0)& \Rightarrow &&\a_9(\b) &= -\b^1+\b^2,&
\nn \a_{10} &=
(1,1,1,0,0,0,0,0,0,0)&\Rightarrow &&\a_{10}(\b) &= \b^1+\b^2+\b^3.&\nonumber
\end{align}
Here the components of each root are the `covariant' coordinates $p_a$
in the $\b^a$ basis of Eq.(\ref{wallform0}) (with $d=10$, and without
dilaton contribution). 
We use the numbering of nodes
and simple roots indicated in fig.~\ref{e10dynk}. In
\cite{DaNi05},  it was  demanded that any (leading) correction be
associated with a root. Since there is no dilaton, this amounts to
requiring $- (N-1) \s$ to be a root.  This occurs only when
$N-1$ is an integer multiple of 3.  The smallest value, $N-1 = 3$,
yields the imaginary, negative root $- 3 \s$, which has length
squared $-10$.

Now, the crucial observation made here is that $-3 \s$ corresponds to
the non-gravity fundamental weight  $\L_{10}$
\be\label{mwts}
\L_{10} = -(3,3,3,3,3,3,3,3,3,3) \quad \Rightarrow\quad
  \L_{10}(\b) =-3\s
\ee
conjugate to $\alpha_{10}$ by (\ref{funwts}). Therefore,
asking that the higher curvature corrections be associated
with dominant non-gravity weights, i.e. of the type $a\L_{10}$ for
some non-negative integer $a$, reproduces the result of \cite{DaNi05} for the
$\E$ prediction of curvature corrections to $D=11$
supergravity. Indeed, equating $a\L_{10}=-3a\s$ to
$-(N-1)\s$ from (\ref{wallform}) yields $N= 3 a + 1$, i.e.
 allowed curvature corrections $R^{1+3a}$, that is $R^4, R^7,
R^{10},\ldots$.

\subsection{Subleading terms and representations}
The pattern found in \cite{DaNi05} for the subleading terms in the
$R^4$ supersymmetry multiplet is
also reminiscent of lowest weight (non-integrable) representations
since the weights associated with these subleading terms (which
are, in the $\E$ case, all on the root lattice) were found to be obtained by
{\em adding positive roots} to the dominant weight $ \Lambda_{10}$.  This
weight pattern does correspond to (the beginning of) a
representation of $\E$, with $ \Lambda_{10}$ as  lowest
weight. It is easily checked that the wall forms corresponding to the
supersymmetry multiplets of the higher-order curvature corrections
$R^{1+3a}$ will similarly resemble the weight patterns of
representations with lowest weight $a \Lambda_{10}$.

\begin{figure} [t!]
\centering 
\includegraphics[scale=0.4]{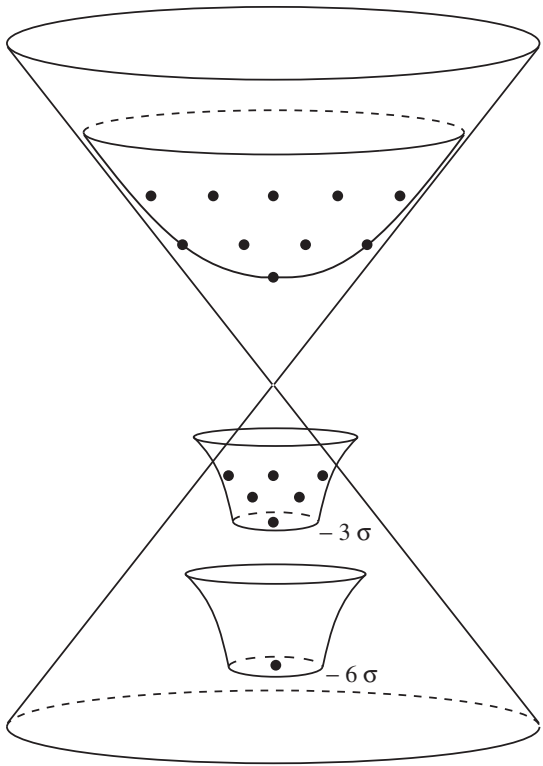}
\caption{\label{rep}\sl
Comparison between the weight pattern of a typical
integrable lowest-weight representation (within the
future light cone), and the weight patterns corresponding
to the supersymmetry multiplets of the curvature corrections
in $M-$theory. The  weight pattern corresponding to $R^{1+3a}$
extends upward from the lowest weight $a \Lambda_{10} = - 3 a \sigma$
which lies within the past light cone.}
\end{figure}

However, these representations are not the usually
considered `integrable lowest-weight' representations \cite{Kac}.
Indeed, integrable lowest-weight representations must have
the {\em negative} of dominant weights as lowest weight. This
is illustrated in figure \ref{rep} .

With the convention that simple
roots are in the {\em future} of the basic spacelike hyperplane
separating positive from negative roots, the fundamental weights
are located in the {\em past} light cone. A typical integrable
lowest-weight diagram would then lie within the future light cone
and would extend {\em upwards} from the negative of a dominant weight
\footnote{Similarly, a highest weight representation would extend
downwards from a dominant weight in the past lightcone.} (see upper
part of figure \ref{rep}). By contrast, the formal weight pattern
corresponding to the curvature correction $R^{1+3a}$ extends upwards
from $a \Lambda_{10}$, which lies in the middle of the past light cone.
Therefore, if these weight patterns do correspond to lowest-weight
representations, these must be of the non-integrable type (which
has not been thoroughly studied because of its greater
mathematical complexity). [We have not checked whether
multiplicities match for the weights that are actually present.]
It is interesting to point out that among the subleading terms
analysed in \cite{DaNi05}, some correspond to weights $\L_{10} +
\a$ involving positive imaginary roots $\a$ up to the $A_9$ level $8$.

It is also interesting to remark that there is another similarity between
the pattern of weights entering curvature corrections, and the weight
diagrams of integrable lowest-weight representations. The weight
diagrams $\left\{ \L \right\} $ of integrable lowest-weight representations 
can be shown \cite{Kac} to be contained in the
convex hull of the quadric $\L^2 =  \L_{lowest}^2$ passing by the
lowest weight $\L_{lowest}$. Rather similarly, it was `botanically'
observed in \cite{DaNi05} that all the wall forms in the $R^4$
supermultiplet considered there are contained within a quadric defined
by the equation $(\L - \L_{lowest})^2 = 2 $. Another observation is
that the enveloping hyperboloid becomes broader as one moves deeper
into the past light cone. For $\L_{lowest}=2\L_{10}$ the condition is
$(\L-\L_{lowest})^2\leq 8$. These facts are illustrated in in figure
\ref{rep}.

\section{Correction terms for other rank~10 KMAs}
\label{DBE10}

We shall now investigate how the weight structure fits with the
other over-extensions which have, in general, more than one
non-gravity root.

\subsection{$DE_{10}$}

We start with $D\E\equiv D_8^{++}$.  This rank 10 hyperbolic KMA is
associated with the bosonic part of pure type I supergravity
\cite{DaHe01,Ju85}.  The corresponding string theory (type I') is
obtained from type I string models by dropping the vector multiplet
and keeping only the gravity multiplet (through a positive charge
orientifold plane).  Note that the bosonic sector is also identical
with the low energy effective action of the closed bosonic string in
ten space-time dimensions. Because $D\E$ is hyperbolic, its
fundamental weights are within or on the {\em past} lightcone. Thus,
the dominant weights that are also roots are necessarily negative,
imaginary roots.

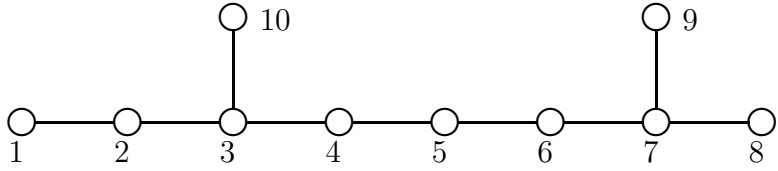
\begin{figure}
\begin{center}
\scalebox{1}{
\begin{picture}(300,60)
\put(5,-5){$1$} \put(45,-5){$2$} \put(85,-5){$3$}
\put(125,-5){$4$} \put(165,-5){$5$} \put(205,-5){$6$}
\put(245,-5){$7$} \put(285,-5){$8$} \put(260,45){$9$}
\put(100,45){$10$} \thicklines
\multiput(10,10)(40,0){8}{\circle{10}}
\multiput(15,10)(40,0){7}{\line(1,0){30}}
\put(250,50){\circle{10}} \put(250,15){\line(0,1){30}}
\put(90,50){\circle{10}}\put(90,15){\line(0,1){30}}
\end{picture}}
\caption{\label{de10dynk}\sl Dynkin diagram of $DE_{10}$ with
numbering
  of nodes.}
\end{center}
\end{figure}

There is one dilaton and the wall forms in this case are given by
\beq\label{simplerootsI} 
\a_1 &=& (0,0,0,0,0,0,0,-1,1;0) \nn 
\vdots\nn
\a_8 &=& (-1,1,0,0,0,0,0,0,0;0) \nn 
\a_9 &=& (1,1,0,0,0,0,0,0,0; -\textstyle{\frac12}) \nn 
\a_{10} &=& (1,1,1,1,1,1,0,0,0;+\textstyle{\frac12}) 
\eeq 
The components listed here are the
`covariant' coordinates $(p_a, p_{\v})$ in the $(\b^a, \v)$ basis
of Eq. (\ref{wallform0}). We use the numbering of nodes indicated
in fig.~\ref{de10dynk}. The first eight nodes give rise to the
symmetry walls (hence they form the gravity line) and the nodes 9
and 10 correspond to the NSNS 2-form and its dual 6-form,
respectively.

The non-gravity fundamental weights are $\L_9$ and $\L_{10}$;
explicitly \be \L_9 (\b,\v)  = - \s -\frac34\v\quad,\quad\quad\quad
\L_{10}(\b,\v) = -\s +\frac14\v. \ee Being a sublattice of $Q(\E)$
($D\E$ is a subalgebra of $\E$ \cite{KlNi04a}), the root lattice
$Q(DE_{10})$ is not self-dual\footnote{The well-known link between
the 11-dimensional metric and 3-form of M-theory on the one hand and
the type-IIA 10-dimensional metric,  dilaton and $p$-forms on the
other hand, dictates the embedding used here. By comparing
Eq.(\ref{simplerootsI}) with Eq.(\ref{simplerootsIIA}) below, one
sees that the simple roots of $D\E$ are given in terms of the simple
roots of $\E$ through $\alpha_i^{D\E} = \alpha_i^{\E}$ ($i = 1,
\cdots, 8$), $\alpha_9^{D\E} = \alpha_{10}^{\E}$ and
$\alpha_{10}^{D\E} = 2 \alpha_9^{\E} + 2 \alpha_{10}^{\E} + 3
\alpha_8 + 4 \alpha_7 + 3 \alpha_6 + 2 \alpha_5 + \alpha_4$.}. In
fact, neither $\L_{9}$ nor $\L_{10}$ is on the root lattice of
$D\E$. This is easy to verify for $\L_{10}$, which has norm squared
equal to $-1$, and can also be checked for $\L_9$ (which is
lightlike). Note that the non-gravity fundamental weight
$\L_{10}^{\E}$ of $\E$  is a positive linear combination of the
non-gravity fundamental weights of $D\E$, \be \L_{10}^{\E} = \L_9 +
2 \L_{10}. \ee

As explained above, we shall explore here the generalized conjecture that
curvature corrections correspond to positive integer combinations
of non-gravity fundamental weights. Let us then consider, within
the context of $DE_{10}$, the dominant weight $a \L_9 + b\L_{10}$,
with $a,b$ some non-negative integers.  Equating this form with a
putative curvature form of the type of (\ref{wallform}) gives the
Einstein frame result
\be
N= 1+a+b,\quad\,\quad\quad\quad K = \frac32 a-\frac12 b.
\ee
In terms of string frame variables (see (\ref{eintostring}))
we obtain
\be\label{pureI}
N = 1+a+b,\quad\,\quad\quad\quad \Ks = -2+2a.
\ee
This result has the following properties
\begin{itemize}
\item{The dilaton coefficients in the string frame which are
compatible with $DE_{10}$ are those which appear in the
genus expansion of closed string perturbation theory, with the
coefficient $a$
of $\L_9$ `counting' the number of string loops.
 We do not have an
understanding of why the ten-dimensional hyperbolic algebra encodes
this `stringy' property outside of supergravity.  We will see this
rather tantalizing property also occurs for the Kac--Moody
correspondants of
some of the other ten-dimensional theories to be studied below.}
\item{For fixed power $N$ of the curvature correction $R^N$ only a
finite number of values for $a,b$ are allowed (since we assumed that
$a\L_9+b\L_{10}$ is dominant, i.e. $a,b\ge 0$). In view of the
string loop interpretation above, this means that only contributions
to $R^N$ from string diagrams with at most $N-1$ loops are
consistent with the structure of the $DE_{10}$ weight lattice.}
\item{The first KM-allowed curvature corrections are at order $R^2$ and arise
when the pair $(a,b)$ takes the values $(0,1)$ or $(1,0)$. Neither of
these weights is a root.  The accompanying power of the dilaton in the
string frame is $e^{-2\P}$ in the first case and $e^{0\P}$ in the
second case suggesting an interpretation as string tree level and
string one loop contribution, respectively.}
\item{Another curious observation is that the known `string tree
level' $R^2$ correction term does not receive a contribution from
$\L_9$ suggesting that in fact there are correction terms related to
it by $SO(9,9)$ rotations. In other words there is an (expected)
sign of T-duality invariance for the corrections. Results in this
direction have been obtained in \cite{KaMe97}.}
\end{itemize}

Because the quantum corrections to type I' string theory have not
been explicitly computed, there is not much to be added here.  Our
above comments are thus predictions on which terms in type I' are
forbidden in the sense of not being compatible with the weight
structure.

\subsection{$BE_{10}$}

\begin{figure}
\begin{center}
\scalebox{1}{
\begin{picture}(340,60)
\put(5,-5){$1$}
\put(45,-5){$2$}
\put(85,-5){$3$}
\put(125,-5){$4$}
\put(165,-5){$5$}
\put(205,-5){$6$}
\put(245,-5){$7$}
\put(285,-5){$8$}
\put(325,-5){$9$}
\put(100,45){$10$}
\thicklines
\multiput(10,10)(40,0){9}{\circle{10}}
\multiput(15,10)(40,0){7}{\line(1,0){30}}
\put(90,50){\circle{10}} \put(90,15){\line(0,1){30}}
\put(290,5){\line(1,0){40}}
\put(290,15){\line(1,0){40}}
\put(310,20){\line(1,-1){10}}
\put(310,0){\line(1,1){10}}
\end{picture}}
\caption{\label{be10dynk}\sl Dynkin diagram of $BE_{10}$ with
numbering of nodes.}
\end{center}
\end{figure}
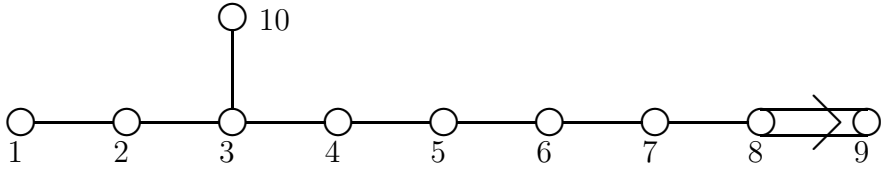

Adding one Maxwell vector multiplet to pure type I supergravity
yields the hyperbolic algebra $BE_{10}\equiv B_8^{++}$
\cite{DaHe01}. Accordingly,
we have the
embeddings $DE_{10}\subset BE_{10}$ and $Q(DE_{10})\subset
Q(BE_{10})$, which is explicitly displayed by observing that
$\a_9^{D\E} = 2 \a_9^{B\E} + \a_8$. The dominant walls are taken
to be 
\beq\label{simplerootsmaxI} 
\a_1 &=& (0,0,0,0,0,0,0,-1,1;0) \nn 
\vdots\nn
\a_8 &=& (-1,1,0,0,0,0,0,0,0;0) \nn 
\a_9 &=& (1,0,0,0,0,0,0,0,0; -\textstyle{\frac14}) \nn 
\a_{10} &=& (1,1,1,1,1,1,0,0,0;+\textstyle{\frac12}) 
\eeq 
We use the numbering of nodes indicated
in fig.~\ref{be10dynk}, with the gravity line consisting of nodes
1 through to 8. Node 9 gives rise to the wall of the Maxwell vector
field and node 10 to the 6-form dual to the NSNS 2-form.

The non-gravity fundamental weights again are $\L_9$ and $\L_{10}$
which are computed to be 
\be 
\L_9(\b,\v) = -\s -\frac34 \v\quad,\quad\quad\quad \L_{10} (\b,\v) =
-\s +\frac14\v. 
\ee 
They 
coincide with the fundamental non-gravity weights of $D\E$.  Note
that $\L_{10}$ is on the root lattice of $B\E$ --- in fact it is a
root, but $\L_9$ is not.  Because $\L_9$ and $\L_{10}$ are the same as
for $D\E$, we again get 
\be 
N = 1+a +b\quad,\quad\quad\quad K= \frac32 a-\frac12 b 
\ee 
in the Einstein
frame. Therefore the $BE_{10}$ compatible string frame corrections
satisfy 
\be\label{einstmax} 
N = 1+a +b\quad,\quad\quad\quad \Ks = -2+2a. 
\ee 
As in the pure supergravity $DE_{10}$ case, the dilaton
coefficients agree with (closed) string perturbation theory.
Eq.~({\ref{einstmax}) also predicts that for $R^N$ only
contributions from string diagrams with up to $(N-1)$ loops are
compatible with the $BE_{10}$ weight lattice structure.

The dominant weight $a \L_9 + b \L_{10}$ is a root of $B\E$ only
for even (non negative) $a$'s.  If one were to restrict the
quantum corrections to be roots, one would miss the corrections
involving an odd number of loops.

\subsection{Heterotic and Type I cases}

\subsubsection{General Considerations}
If one adds $k$ abelian vector multiplets to pure type I supergravity in
ten dimensions, the relevant Kac--Moody algebra is
$\mathfrak{so}(8,8+k)^{++}$ \cite{HeJu03}, which is a non-split
real form.   The analysis proceeds in this case in a way very
similar to that of $B\E$, because it is that subalgebra that
controls the (real) roots and weights.

To understand this point, consider first the familiar toroidal
dimensional reduction to three spacetime dimensions.  After
dualization of all non-metric fields to scalars, the theory is
described by the three-dimensional Einstein-Hilbert action coupled
to the non-linear sigma model action for the coset space
$SO(8,8+k)/SO(8)\times SO(8+k)$.  This action is most easily
written down using the Iwasawa decomposition appropriate to
non-split real forms \cite{Helgason}.  This decomposition, in
turn, follows from the Tits-Satake decomposition of the algebra
$\mathfrak{so}(8,8+k)$ in terms of real root spaces, which we
briefly recall.  The split algebra $B_8 \equiv \mathfrak{so}(8,9)$
is a maximal split subalgebra of $\mathfrak{so}(8,8+k)$.   One can
decompose over the reals $\mathfrak{so}(8,8+k)$ in terms of
representations of $\mathfrak{so}(8,9)$.  The Cartan generators of
$B_8$ have indeed real eigenvalues, while the other (compact)
Cartan generators of $\mathfrak{so}(8,8+k)$ have imaginary
eigenvalues and cannot be diagonalized over the reals.  Because
the rank of $B_8$ (known also as the real rank of
$\mathfrak{so}(8,8+k)$) is eight, the weights are 8-dimensional
vectors. They turn out to coincide with the roots of $B_8$, i.e.,
with the weights of the adjoint representation of $B_8$, but they
come with a non trivial multiplicity. Specifically, the short
$B_8$-roots appear $k$ times, while the long roots are
non-degenerate.  And furthermore, the zero eigenvalue also appears
in the spectrum. Their associated eigenvectors are the elements of
$\mathfrak{so}(k)$.  Corresponding to this decomposition, the
coset Lagrangian for $SO(8,8+k)/SO(8)\times SO(8+k)$ takes a form
very similar to that of $SO(8,9)/SO(8)\times SO(9)$, namely: ($i$)
there are 8 dilatons with standard kinetic term, because 8 is the
real rank of $\mathfrak{so}(8,8+k)$ i.e., the rank of its maximal
split subalgebra $\mathfrak{so}(8,9)$; ($ii$) for each positive root
of $B_8$, {\em counting $\mathfrak{so}(8,8+k)$ multiplicities},
there is an axion with a kinetic term multiplied by the
exponential of the corresponding root.

Coming back to the non reduced model, it is natural to conjecture
that it is dual to the geodesic motion on the infinite dimensional
coset space $SO(8,8+k)^{++}/ K(SO(8,8+k)^{++})$.  Again,  the algebra
$\mathfrak{so}(8,8+k)^{++}$ is a representation of its maximal
split subalgebra  $\mathfrak{so}(8,9)^{++} \equiv B_8^{++} \equiv
B\E$.    The `real roots' of $\mathfrak{so}(8,8+k)^{++}$ are the
weights of that representation.  They are equal to the roots of $
B\E$, but come with some non trivial multiplicity (over and above
the Kac--Moody multiplicity of the $B\E$ imaginary roots).  For
instance, the real root $\a_9$ has multiplicity $k$.  It is the
maximally split subalgebra $B\E$ that one sees in the $\sigma$
model Lagrangian.  In particular, the billiard region is the same
as in the case of one Maxwell multiplet, the only difference being
that the electric wall (associated with the simple root $\a_9$)
appears $k$ times.

As the representations of $\mathfrak{so}(8,8+k)^{++}$ are
characterized by real weights that are weights of $B_8^{++} \equiv
B\E$, the quantum corrections should be associated with weights of
$B\E$.  Note that the non-trivial multiplicities related to the
fact that the algebra is a non-split real form concerns only the
terms involving the gauge fields, {\em which are subleading}.

$\mathfrak{so}(8,8+k)^{++}$-supergravity theories, i.e.,
supergravity theories with 16 supercharges, one gravity multiplet
and $k$ vector multiplets, have been analyzed in the string
context in \cite{Keur}. It was found that string backgrounds
consistent with this matter content could exist, but only in
particular spacetime dimensions, which depend on $k$.  For
instance, the $k=1$ theory has global anomalies except in 3
spacetime dimensions and below.  For $k=2$, the maximal spacetime
dimension is 5. This suggests that it would be of interest to
repeat the $B\E$ weight analysis in lower dimensions. However,
since the effective actions for those theories have not been much
studied, we have not performed here that analysis.

If one wants to generalize the abelian group $U(1)^k$ to some
non-abelian gauge group (as required for the heterotic string),
one encounters the difficulty that there is no `nice' and obvious
choice of KMA that would naturally accommodate the Yang Mills
gauge groups.  Because the rank of the gauge groups relevant to
string theory in 10 dimensions is $16$, one might nevertheless
argue that it is the algebra $\mathfrak{so}(8,8+16)^{++} =
\mathfrak{so}(8,24)^{++}$ and thus the maximal split algebra $B\E$
that controls the weight pattern.  Evidence for this comes from
the billiard analysis \cite{DaHe01} and was also given in
\cite{BrGaGaHe05}.  The above study of quantum corrections for
$B\E$  would therefore again apply.\footnote{There is a fourth
hyperbolic rank 10 Kac--Moody algebra, namely $C\E$.  This algebra
is dual to $B\E$ and  is a twisted overextension \cite{HeJu03},
although $C\E\neq C_8^{++}$. Because it has not been associated to
a field theoretical model, we shall not investigate it here.}

Because quantum corrections to the heterotic string and type I
string models have been studied, one can check whether they are
compatible with the $BE_{10}$ weight structure.  The analysis
proceeds differently in the two cases.

\subsubsection{Heterotic String}

The first known curvature corrections to the heterotic string are
at order $R^2$ and by studying eq.~(\ref{pureI}) we see that there
are two solutions compatible with $BE_{10}$ when the pair $(a,b)$
takes the values $(0,1)$ or $(1,0)$. The accompanying power of the
dilaton in string frame is $e^{-2\P}$ in the first case and
$e^{0\P}$ in the second case corresponding to string tree level
and string one loop contribution. In fact the first term (tree
level) has been computed in \cite{GrSl86} and agrees with our
result. The second term (one-loop), however, has been argued to be
absent \cite{Tseytlin}.  This shows that the algebraic constraints
heuristically investigated here may play
 a role somewhat analogous to selection rules: they can be
 used to predict which terms should be absent, but not which
 terms should be actually present (indeed, further hidden
 symmetries might cancel a term allowed by  general
  selection rules).  The same situation holds for $R^3$
corrections, which are allowed at tree level, one-loop and two-loops
by our algebraic constraints, but which have been argued to be
absent on account of supersymmetry \cite{Tseytlin} (see also
\cite{Metsaev:1986yb,Bergshoeff:1989de}). Finally, $R^4$
corrections up to three loops are permitted by the $B\E$ algebraic
constraints. However, although $R^4$ corrections are known to be
present in the heterotic effective action, their actual loop
dependence is less clear \cite{Tseytlin}.

\subsubsection{Type I} To analyze the type I superstring,
one must recall that the transition from the Einstein frame (where
we have derived the algebraically compatible counterterms) to the
string frame is different than in the heterotic case: the type I
dilaton is minus the heterotic dilaton, and the spacetime metric
changes accordingly.  Converting the algebraically compatible
counterterms to the type I string frame, one finds that the two
$R^2$ corrections found above come with the (type I) dilaton powers
$\exp(- \P)$ and $\exp(-3 \P)$, respectively.  The first term
corresponds to the tree heterotic correction, while the second
corresponds to the heterotic one-loop term.  Only the first term is
compatible with the type I effective action and furthermore, it is
in perfect agreement with \cite{Tseytlin}.  Combining the
information that the second term must be absent in type I with
heterotic-type I duality, one can argue that there is no one-loop
$R^2$ term in the heterotic case.  The same argument is too weak to
eliminate all $R^3$ corrections, which are forbidden by
supersymmetry.

\section{Other algebras}
\label{g++}

\subsection{Results}
For the other over-extended algebras we give the result only in
tabulated form, including also the results obtained for the rank $10$
hyperbolic KMAs in the preceding section for completeness.
We take the
standard gravity lines (see for example \cite{DadeBuHeScho02}) and
consider positive integral linear combinations of the non-gravity
fundamental weights and equate them to the general leading correction
wall form (\ref{wallform}). We exclude $C_n^{++}$ from the list since
the corresponding $(3+1)$-dimensional theory has $(n-3)$ dilaton
fields which would clutter the notation. It can be checked however
that the algebra allows for curvature correction $R^N$ for all values
of $N$. \\

\begin{center}
\begin{tabular}{|c|c|c|c|}
\hline
Algebra & $(d+1)$ & $N$ in $\sqrt{-G}\, R^N e^{K\v}$ & $K$
in $\sqrt{-G}\, R^N e^{K\v}$ \\
\hline
\hline
$A_n^{++}$ & $n+3$ & $1+a$ & --\\
$B_n^{++}$ & $n+2$ & $1+a+b$ & $\frac{n-2}{\sqrt{2n}}a
   -\frac2{\sqrt{2n}}b$ \\
$D_n^{++}$ & $n+2$ & $1+a+b$ & $\frac{n-2}{\sqrt{2n}}a
   -\frac2{\sqrt{2n}}b$ \\
$G_2^{++}$ & $5$ & $1+a$ & -- \\
$F_4^{++}$ & $6$ & $1+2a+b$ & $-\frac1{\sqrt{2}}b$ \\
$E_6^{++}$ & $8$ & $1+2a+b$ & $-b$ \\
$E_7^{++}$ & $9$ & $1+a+2b$ &
$\frac3{\sqrt{7}}a-\frac1{\sqrt{7}}b$ \\
& $10$ & $1+2a$ & -- \\
$E_8^{++}$ & $11$ & $1+3a$ & -- \\
\hline
\end{tabular}
\end{center}

As for the just quoted $C_n^{++}$ case, one sees that all the
algebras in the table (except $E_8^{++}$ and $E_7^{++}$ in ten
dimensions) are compatible with curvature corrections $R^N$ for all
positive integers $N$. This is not very restrictive but the fact
that only integers $N$ (as opposed to non integer values) are forced
by the algebraic requirement was not a priori guaranteed.
Furthermore, many of the corrections known to be present in the
effective Lagrangians would not be allowed had we insisted on having
only roots.  For instance, for pure gravity in spacetime dimension
$D=d+1$, the corrections corresponding to roots are of the restricted
form $R^{k(D-2)+1}$.

The $\E$ case corresponds to M-theory and has been analysed above.
The entry with $D=10$ for $E_7^{++}$ in the table corresponds to a
theory which is a non-supersymmetric truncation of IIB
supergravity, keeping only gravity and the four-form potential.
Therefore one has the same problem with writing a manifestly
covariant Lagrangian for this theory in ten dimensions as with IIB
supergravity and the theory is usually considered in $D=9$, where
all powers of the curvature are allowed.  Here we see however that
corrections in ten dimensions would be compatible with the algebra
structure only for odd powers of the curvature.

In the case where there is a dilaton, non trivial restrictions on
the allowed powers of the dilaton for a given $N$ do appear, as
indicated in the right column of the table.  However, it is
difficult to test the validity of these predictions as the quantum
corrections appearing in the effective Lagrangians of the
corresponding theories have not been investigated.

\subsection{String frame}
The dilaton entries $K$ in the table are in the Einstein frame.
The notion of string frame really makes sense only for the
string-related theories, but one can nevertheless explore the
consequences of converting the formulas  into a would-be string
frame by using Eq.~(\ref{eintostring1}). The result for $B_n^{++}$
and $D_n^{++}$ is \be \Ks = -2+ 2 \, a \quad\quad\quad
   (B_n^{++}\,{\rm{ and }}\, D_n^{++}),
\ee Thus, $\Ks$ is an even integer that depends only on $a$
suggesting an interpretation in terms of a genus expansion in
string perturbation theory for all $n$.

For the three exceptional algebras which give rise to a dilaton one
finds after conversion to the string frame
\begin{align}
\Ks &= -2+2\, a  &\quad\quad (F_4^{++})&,\\
\Ks &= -2+\frac43a + \left( \frac23- \frac{2}{\sqrt{3}}
\right) b &\quad\quad (E_6^{++})&,\\
\Ks &= -2+ \left(4 + 6 \sqrt{2}\right)\frac{a}{7}
   +\left(8 - 2 \sqrt{2}\right)\frac{b}{7}
   &\quad\quad (E_7^{++})&.
\end{align}
$\Ks$ for $F_4^{++}$ is identical to $\Ks$ for $B_n^{++}$ and
$D_n^{++}$. For the other two cases, we do not have a transparent
interpretation.

\section{Reduction of counterterms, IIA and IIB}
\label{EAB}

It is conjectured that $E_{10}\equiv E_8^{++}$ is also the relevant
symmetry for type IIA and type IIB supergravity in ten dimensions.
Evidence for this conjecture was given in \cite{DaHe01,KlNi04a,KlNi04b},
following earlier work on the embedding of these theories into $E_{11}$
\cite{We01,SchnWe01,SchnWe02,We04}. The structure of the low derivative
curvature correction terms for these theories is known from string
scattering computations and is summarized for example in
\cite{PeVaWe01}. Somewhat unexpectedly, in both cases there is strong
evidence that the first correction appears at order $R^4$ and receives
contributions only from string tree level and string one-loop
diagrams.\footnote{To be sure, there are corrections coming from $D(-1)$
  instantons for IIB which we do not discuss here. The absence of
  higher order loop corrections was partially confirmed in an explicit
  computation  \cite{Ie02}.}
We will now examine these results in the light of
our algebraic compatibility conditions.

\subsection{IIA}

The IIA supergravity theory can be obtained by standard dimensional
reduction of the $D=11$ theory. Reducing to Einstein frame we find
that the dominant walls (simple roots) of $E_{10}$ are now expressed
via
\beq\label{simplerootsIIA}
\a_1 &=& (0,0,0,0,0,0,0,-1,1;0) \nn
\vdots \nn
\a_8 &=& (-1,1,0,0,0,0,0,0,0;0) \nn
\a_9 &=& (1,0,0,0,0,0,0,0,0; + \textstyle{\frac34}) \nn
\a_{10} &=& (1,1,0,0,0,0,0,0,0; -\textstyle{\frac12})
\eeq
Here the first eight `symmetry roots' define the new (non-maximal)
gravity line of the $\E$ Dynkin diagram (see fig.~\ref{e10dynk}),
corresponding to the symmetry
walls of gravity in $D=10$ space-time dimensions. Unlike for $D=11$
supergravity, the simple root $\a_9$ is no longer associated with a
symmetry wall, but now corresponds to the Kaluza Klein vector; the
simple root $\a_{10}$ is associated with the NSNS two-form.

The relevant fundamental weights not belonging to the gravity line are
now
\beq\label{iiawts}
\L_9 &=& - (2,2,2,2,2,2,2,2,2;+ \textstyle{\frac12}) \quad\Rightarrow\quad
\L_9(\b,\v) =-2\s +\frac12\v\nn
\L_{10} &=& - (3,3,3,3,3,3,3,3,3; -\textstyle{\frac14})
 \quad\Rightarrow\quad \L_{10}(\b,\v) = -3\s-\frac14\v
\eeq
Equating the dominant weight $a\L_9 + b \L_{10}$ with
(\ref{wallform}) corresponds to the higher order term \be
\sqrt{-G}\,R^N e^{K\v} \quad \mbox{with} \quad N= 1+ 2a + 3b \; ,
\;\; K = - a + \frac{b}2. \ee In string frame the dilaton exponent
gives \be \Ks = -2 + 2b \ee according to (\ref{eintostring}), so
that the coefficient $b$ counts the number of string loops. We see
that $\L_{10}$ -- which corresponded to $R^4$ in $D=11$ supergravity
-- in the IIA basis corresponds to $(a,b)=(0,1)$, i.e. $R^4$ at one
loop in string perturbation theory. Our reasoning allows for tree
level terms ($b=0$) for all odd powers of the curvature. The absence
of $R^3$ corrections, however, has been established both by
supersymmetry arguments and explicit calculation. If one
considers the known tree level term for $R^4$ one finds a wall
form which is the following sum of simple roots

\be\label{iiatree}
-\left(3\a_1+6\a_2+7\a_3+12\a_4+15\a_5+18\a_6+21\a_7
  +\frac{27}2\a_8+6\a_9+\frac{21}2\a_{10}\right).
 \ee
Because of the appearance of fractional coefficients, this
 is not element of the weight lattice (=root lattice).
Therefore, $\E$
predicts correctly only the maximum loop order at which corrections can
occur. One way to
interpret this apparent discrepancy between the known string
computations and the present KMA analysis is that the Kac--Moody model
of \cite{DaHeNi02} is thought to describe the {\em decompactified}
version of M-theory, which in particular involves taking the limit
in which the string coupling (or equivalently $\P$) tends to infinity.
In this strong coupling limit only the highest genus terms of a string
loop expansion survive. Indeed, only the one loop terms is known to
lift to eleven dimensions.

\subsection{IIB}

One gets stranger results for type IIB string theory.  This is
perhaps not surprising in view of the singular field theory limit
involved in getting to type IIB from the 11-dimensional model.  For
IIB in $D=10$ the $\E$ simple roots are now represented as
\beq\label{simplerootsIIB} 
\a_1 &=& (0,0,0,0,0,0,0,-1,1;0) \nn 
\vdots\nn
\a_7 &=& (0,-1,1,0,0,0,0,0,0;0) \nn 
\a_{10} &=& (-1,1,0,0,0,0,0,0,0;0) \nn 
\a_8 &=& (1,1,0,0,0,0,0,0,0;-\textstyle{\frac12}) \nn 
\a_9 &=& (0,0,0,0,0,0,0,0,0; + 1) 
\eeq 
As before, the first
eight are symmetry roots, but the associated `gravity line' is now
given by $\a_1,\dots,\a_7,\a_{10}$, and thus differs from the IIA
gravity line (compare also fig.~\ref{e10dynk}). The root  $\a_{8}$
corresponds to the wall generated by the NSNS 2-form, while $\a_9$
corresponds to the dilaton wall. Remarkably, and unlike for the IIA
theory, the dilaton root $\a_9$ has no components involving the spatial
neunbein.

Now the relevant fundamental weights are
\beq
\L_8 &=& - (4,4,4,4,4,4,4,4,4;0) \quad\Rightarrow\quad
\L_8(\b,\v) = -4\s \nn
\L_9 &=& - (2,2,2,2,2,2,2,2,2;- \textstyle{\frac12}) \quad\Rightarrow\quad
\L_9(\b,\v) = -2\s + \frac12 \v
\eeq
Demanding that the dominant weight $a\L_8 + b\L_9$ be identical to the
wall (\ref{wallform}) yields corrections in the Einstein frame with
\be
N = 1 + 4a + 2b,\quad\quad K= -b.
\ee
Conversion to the string frame gives
\be\label{IIBK}
N = 1 + 4a + 2b,\quad\quad \Ks = -2 + 2a,
\ee
so that the coefficient $a$ counts the number of string loops.

The first correction terms compatible with this weight pattern are
$R^3$ (tree) and $R^5$ (tree and 1-loop).  The pattern does not
match with the known $\sqrt{-G}\,R^4$ in ten dimensions, independently  of
the dilatonic factor, which would require a wall form  $\propto
(3,3,3,3,3,3,3,3,3;*)$. One can also check directly that the
latter combination does not lie on the $\E$ root lattice, no
matter how the dilaton factor is chosen. Incidentally, the tree
level term $\sqrt{-G}\,R^4 e^{-3\v/2}$ in this case is again
located at the same root vector (\ref{iiatree}) as in the IIA
case. The one loop term is also off the weight lattice for IIB.

The correction term weight $\L_{10}$ which gave sensible results for
$D=11$ and IIA (see (\ref{mwts}) and (\ref{iiawts})) in this basis becomes
\be
\L_{10} = - (4,3,3,3,3,3,3,3,3;0) \quad\Rightarrow\quad
\L_{10}(\b,\v) = -3\s - \b^1
\ee
and is non-isotropic for IIB in $D=10$. This might be related to the
difficulties with
writing a covariant Lagrangian for the IIB supergravity theory in ten
space-time dimensions and we are therefore
tempted to consider the situation after compactification of the
theory on $S^1$ (coordinatized by $x^1$) to nine space-time dimensions
(which is required to make the T-duality equivalence of IIA and IIB manifest),
as is necessary in all string calculations of higher order effects
in IIB theory. The 11-component of the spatial vielbein becomes
another `dilaton', and we would need to match $\sqrt{-G}\,R^4$
only with $\propto (3,3,3,3,3,3,3,3;*;*)$. This is certainly
possible and will be the reduction of the one-loop term of the IIA
theory but only at the price of including a non-covariant term
involving this extra dilatonic factor, so far not seen in string
perturbation theory.

\section{Conclusions}
\label{concl}

In this paper, we have first shed new
light on the findings of \cite{DaNi05} concerning the quantum
corrections of M-theory.  We have emphasized that the imaginary
roots uncovered in \cite{DaNi05} are dominant
non-gravity weights, suggesting a representation
theoretic interpretation for the subleading quantum corrections.

We then `botanically' explored the possible consequences for
various maximally oxidized gravitational theories of a general
conjecture linking dominant non-gravity weights to quantum
curvature corrections. This conjecture has been found to have
quite a few successes, but also a certain number of failures.
Among the successes, let us mention that the consideration of weights
(as opposed to roots) makes a difference for non self-dual lattices,
and has been found to be necessary to reproduce some of the known
quantum corrections. The conjecture derives further credit from
the fact that the relevant KMAs appear to `know' about string
perturbation theory, via eqns.~(\ref{pureI}), (\ref{einstmax}) and
(\ref{IIBK}). On the other hand,  for the type I models, we found that
the algebraic restrictions imposed by
the conjectured symmetry are not restrictive enough:
they allow terms that have not been observed.
In the case of the types IIA and IIB, they also forbid terms
that are known to occur --- independently of whether one
considers roots or weights. From
this point of view, it is for the $\E$-based M-theory that the
matching between the algebraic constraints and the known results
works most successfully (and indeed spectacularly so).

{}For the non string-related theories like pure gravity, the
matching does appear to make sense since the known quantum
corrections are reproduced (when one includes weights), but it is
more difficult to test the Kac--Moody predictions because little is
in general known on the effective lagrangians.  It is expected that
a deeper understanding of the occurrence of Kac--Moody weights could
be obtained through a further analysis of the allowed $\sigma$-model
counterterms, which are controlled by the invariants of the
`maximal compact subgroup' $K(G^{++})$ of $G^{++}$.

\vspace{.3cm}

\noindent {\it Note added:} While this article was completed, the
preprint \cite{Lambert:2006he} was posted. This preprint (whose
reference [21] refers to part of the present work) shows that
reduction to three spacetime dimensions of curvature corrections
involves weights of the duality algebra $G$ which is manifest in
three dimensions --- and of which the Kac--Moody algebra
$G^{++}$ is the overextension.\\

\noindent {\bf Acknowledgements:} We thank V.~Kac, K.~Peeters and
S.~Theisen for informative discussions. We are grateful to
Marie-Claude Vergne for preparing figure 2. AH and MH would like to
thank the KITP of UCSB for kind hospitality during the early stages
of this project.  TD acknowledges the support of the European
Research and Training Network `Forces Universe' (contract number
MRTN-CT-2004-005104). The work of MH is partially supported  by IISN
- Belgium (convention 4.4505.86), by the `Interuniversity Attraction
Poles Programme -- Belgian Science Policy' and by the European
Commission programme MRTN-CT-2004-005104, in which he is associated
to V.U. Brussel.  AK and HN are grateful for the hospitality of ULB
and IHES during early stages of this work and
acknowledge the support of the European
Research and Training Network `Superstrings' (contract number
MRTN-CT-2004-512194). Research also supported in part by the CTP and
the LNS of MIT and the U.S. Department of Energy under cooperative
agreement $\#$DE-FC02-94ER40818.


\baselineskip14pt

\begin{thebibliography}{30}

\bibitem{BKL} V.~A.~Belinsky, I.~M.~Khalatnikov and E.~M.~Lifshitz,
  {\sl Oscillatory Approach To A Singular Point In The Relativistic
  Cosmology},  Adv.\ Phys.\  {\bf 19} (1970) 525

\bibitem{DaHe01}  T.~Damour and M.~Henneaux, {\sl  $E_{10}$, $BE_{10}$
  and arithmetical chaos in superstring cosmology}, Phys.\ Rev.\
  Lett.\  {\bf 86} (2001) 4749, {\tt hep-th/0012172}

\bibitem{DadeBuHeScho02}  T.~Damour, S.~de Buyl, M.~Henneaux and
  C.~Schomblond, {\sl Einstein billiards and overextensions of
  finite-dimensional simple Lie algebras}, JHEP {\bf 0208} (2002) 030,
  {\tt hep-th/0206125}

\bibitem{DaHeNi03}  T.~Damour, M.~Henneaux and H.~Nicolai, {\sl
  Cosmological billiards}, Class.\ Quant.\ Grav.\  {\bf 20} (2003)
  R145, {\tt hep-th/0212256}

\bibitem{DHJN} T.~Damour, M.~Henneaux, B.~Julia and H.~Nicolai,
{\sl Hyperbolic Kac-Moody algebras and chaos in Kaluza-Klein
models}, Phys.\ Lett.\ B {\bf 509}, 323 (2001), {\tt hep-th/0103094}

\bibitem{DaNi05}  T.~Damour and H.~Nicolai, {\sl Higher order M
  theory corrections and the Kac-Moody algebra $E_{10}$},   Class.\
  Quant.\ Grav.\  {\bf 22} (2005) 2849, {\tt hep-th/0504153}

\bibitem{MiMoYa05}  S.~Mizoguchi, K.~Mohri and Y.~Yamada, {\sl
  Five-dimensional supergravity and hyperbolic Kac-Moody algebra
  $G_2^H$}, {\tt hep-th/0512092}

\bibitem{CS} J.H.~Conway and N.J.A.~ Sloane, {\sl Sphere packings,
  lattices and groups}, Grundlehren der mathematischen Wissenschaften,
  Vol. 290, 2nd ed. (Springer verlag, 1991)

\bibitem{CrJuLuPo99} E.~Cremmer, B.~Julia, H.~Lu and C.~N.~Pope, {\sl
  Higher-dimensional origin of D = 3 coset symmetries}, {\tt
  hep-th/9909099}

\bibitem{Kac}
V.G.~Kac, {\it Infinite Dimensional Lie Algebras}, 3rd edn., Cambridge
University Press, 1990

\bibitem{Ju85} B.~Julia, in: Lectures in Applied Mathematics, Vol. 21
  (1985), AMS-SIAM, p. 335; preprint LPTENS 80/16

\bibitem{KlNi04a} A.~Kleinschmidt and H.~Nicolai, {\sl $E_{10}$ and
  $SO(9,9)$ invariant supergravity}, JHEP {\bf 0407} (2004) 041, {\tt
  hep-th/0407101}

\bibitem{KaMe97}  N.~Kaloper and K.~A.~Meissner, {\sl  Duality beyond
  the first loop},  Phys.\ Rev.\ D {\bf 56} (1997) 7940, {\tt
  hep-th/9705193}

\bibitem{HeJu03} M.~Henneaux and B.~Julia,
{\sl Hyperbolic billiards of pure D = 4 supergravities},
 JHEP {\bf 0305}, 047 (2003), {\tt hep-th/0304233}

\bibitem{Helgason} S.~Helgason, {\sl Differential Geometry, Lie
  Groups, and Symmetric Spaces}, Graduate Studies in
  Mathematics, Vol. 34, Amer. Math. Soc. (Providence, 2001)

\bibitem{Keur} A. Hanany, B. Julia and A. Keurentjes, {\sl unpublished};
A.~Keurentjes, {\sl Classifying orientifolds by flat n-gerbes},
JHEP {\bf 0107}, 010 (2001), {\tt hep-th/0106267}


\bibitem{BrGaGaHe05} J.~Brown, S.~Ganguli, O.~J.~Ganor and
  C.~Helfgott, {\sl $E_{10}$ orbifolds}, JHEP {\bf 0506} (2005) 057,
  {\tt hep-th/0409037}


\bibitem{GrSl86} D.~J.~Gross and J.~H.~Sloan, {\sl The Quartic
  Effective Action For The Heterotic String},  Nucl.\ Phys.\ B {\bf
  291} (1987) 41.

\bibitem{Tseytlin} A.~A.~Tseytlin,
{\sl Heterotic - type I superstring duality and low-energy effective
actions}, Nucl.\ Phys.\ B {\bf 467} (1996) 383
{\tt hep-th/9512081}

\bibitem{Metsaev:1986yb}  R.~R.~Metsaev and A.~A.~Tseytlin, {\sl
 Curvature Cubed Terms In String Theory Effective Actions},
  Phys.\ Lett.\ B {\bf 185} (1987) 52

\bibitem{Bergshoeff:1989de}   E.~A.~Bergshoeff and M.~de Roo, {\sl The
  Quartic Effective Action Of The Heterotic String And
  Supersymmetry},  Nucl.\ Phys.\ B {\bf 328} (1989) 439.

\bibitem{KlNi04b}  A.~Kleinschmidt and H.~Nicolai, {\sl IIB
supergravity and $E_{10}$}, Phys.\ Lett.\ B {\bf 606} (2005) 391,
{\tt hep-th/0411225}

\bibitem{We01} P.~C.~West, {\sl $E_{11}$ and M theory}, Class.\
  Quant.\ Grav.\  {\bf 18}, 4443 (2001), {\tt hep-th/0104081}

\bibitem{SchnWe01} I.~Schnakenburg and P.~C.~West, {\sl Kac--Moody
  symmetries of 2B supergravity}, Phys. Lett. B {\bf 517} (2001)
  421--428, {\tt hep-th/0107081}

\bibitem{SchnWe02} I.~Schnakenburg and P.~C.~West, {\sl Massive
  IIA supergravity as a nonlinear realization}, Phys. Lett. B
  {\bf 540} (2002) 137--145, {\tt hep-th/0204207}

\bibitem{We04}  P.C.~West, {\sl The IIA, IIB and eleven dimensional
  theories and their common $E_{11}$ origin}, Nucl.\ Phys.\ B {\bf
  693}, 76 (2004), {\tt hep-th/0402140}

\bibitem{PeVaWe01}  K.~Peeters, P.~Vanhove and A.~Westerberg, {\sl
  Supersymmetric higher-derivative actions in ten and eleven
  dimensions,  the associated superalgebras and their formulation in
  superspace}, Class.\ Quant.\ Grav.\  {\bf 18} (2001) 843, {\tt
  hep-th/0010167}

\bibitem{Ie02}  R.~Iengo, {\sl Computing the $R^4$ term at two
  super-string loops},   JHEP {\bf 0202} (2002) 035, {\tt
  hep-th/0202058}

\bibitem{DaHeNi02} T.~Damour, M.~Henneaux and H.~Nicolai, {\sl
    $E_{10}$ and a "small tension expansion" of M-theory}, Phys.
    Rev. Lett. {\bf 89} (2002) 221601, {\tt hep-th/0207267}

\bibitem{Lambert:2006he}
N.~Lambert and P.C.~West, {\sl Enhanced Coset Symmetries and Higher
Derivative Corrections}, {\tt hep-th/0603255}


\end{thebibliography}
\end{document}